\documentclass[11pt]{article}
\usepackage{moriond,epsfig}

\newcommand{\newc}{\newcommand}
\newc\eg{{\it {e.g.}}}  \newc\etal{{\it {et al.}}} \newc\ie{{\it i.e.}}
\newc\etc{{\it {etc}}}  

\newcommand\lsim{\mathrel{\rlap{\lower4pt\hbox{\hskip1pt$\sim$}}
    \raise1pt\hbox{$<$}}}
\newcommand\gsim{\mathrel{\rlap{\lower4pt\hbox{\hskip1pt$\sim$}}
    \raise1pt\hbox{$>$}}}

\newc{\mchi}{m_\chi}
\newc{\charone}{\chi_1^\pm} \newc{\mcharone}{m_{\chi_1^\pm}}

\newc{\hl}{h}               \newc{\mhl}{m_{\hl}}
\newc{\hh}{H}               \newc{\mhh}{m_{\hh}}
\newc{\ha}{A}               \newc{\mha}{m_{\ha}}
\newc{\hc}{H^{\pm}}         \newc{\mhc}{m_{\hc}}

\newc{\qzero}{Q_0}          \newc{\qstop}{Q_{\widetilde t}}

\newc{\amu}{a_{\mu}}        \newc{\amususy}{a_{\mu}^{\rm SUSY}}
\newc{\amuexpt}{a_{\mu}^{\rm expt}}        \newc{\amusm}{a_{\mu}^{\rm SM}}
\newc{\deltaamususy}{\Delta a_{\mu}^{\rm SUSY}}

\newc{\msbar}{\overline {\rm MS}} \newc{\drbar}{\overline {\rm DR}}

\newc{\mt}{m_t} \newc{\mb}{m_b} \newc{\mtau}{m_{\tau}}
\newc{\yt}{h_t} \newc{\yb}{h_b} \newc{\ytau}{h_{\tau}}
\newc{\mtpole}{m_t^{\rm pole}} \newc{\mbpole}{m_b^{\rm pole}} 
\newc{\mtaupole}{m_{\tau}^{\rm pole}} 

\newc{\mtmtsmmsbar}{m_t(m_t)^{\msbar}_{{\rm SM}}}
\newc{\mtmtsmdrbar}{m_t(m_t)^{\drbar}_{{\rm SM}}}
\newc{\mtmtmssmdrbar}{m_t(m_t)^{\drbar}_{{\rm SUSY}}}

\newc{\mbmbsmmsbar}{m_b(m_b)^{\msbar}_{{\rm SM}}}
\newc{\mbmzsmmsbar}{m_b(\mz)^{\msbar}_{{\rm SM}}}
\newc{\mbmzsmdrbar}{m_b(\mz)^{\drbar}_{{\rm SM}}}
\newc{\mbmzmssmdrbar}{m_b(\mz)^{\drbar}_{{\rm SUSY}}}
\newc{\mtaumzsmmsbar}{m_{\tau}(\mz)^{\msbar}_{{\rm SM}}}
\newc{\mtaumzsmdrbar}{m_{\tau}(\mz)^{\drbar}_{{\rm SM}}}
\newc{\mtaumzmssmdrbar}{m_{\tau}(\mz)^{\drbar}_{{\rm SUSY}}}

\newc{\mgut}{M_{\rm GUT}}

\newc{\mhalf}{m_{1/2}}      \newc{\mzero}{m_0}

\newc{\tanb}{\tan\beta}
\newc{\azero}{A_0}
\newc{\at}{A_t} \newc{\abot}{A_b} \newc{\atau}{A_\tau} 
\newc{\bmu}{B\mu}           \newc{\sgn}{{\rm sgn}}
\newc{\mone}{M_1}           \newc{\mtwo}{M_2}

\newc\br{\mbox{BR}}
\newc{\abundchi}{\Omega_\chi h^2}
\newcommand\tev{\,\mbox{TeV}}
\newcommand\gev{\,\mbox{GeV}}

\newc{\ra}{\rightarrow}

\newc{\beq}{\begin{equation}}
\newc{\eeq}{\end{equation}}
\newc{\bea}{\begin{eqnarray}}
\newc{\eea}{\end{eqnarray}}
\newc\alphas{\alpha_s}

\newc\alphaem{\alpha_{em}}

\newcommand\mz{m_{Z}}

\newc\supl{{\widetilde u}_L}      \newc\msupl{m_{\supl}}
\newc\supr{{\widetilde u}_R}      \newc\msupr{m_{\supr}}
\newc\sdl{{\widetilde d}_L}      \newc\msdl{m_{\sdl}}
\newc\sdr{{\widetilde d}_R}      \newc\msdr{m_{\sdr}}

\newcommand\stopone{{\widetilde t}_1}   \newcommand\mstopone{m_{\stopone}}
\newcommand\stoptwo{{\widetilde t}_2}   \newcommand\mstoptwo{m_{\stoptwo}}

\newcommand\stauone{{\widetilde \tau}_1}

\newc\hpm{H^\pm} \newc\hp{H^+} \newc\hm{H^-} 
\newc\sfermion{\tilde f}  \newc\msfermion{m_{\sfermion}}  

\def\Journal#1#2#3#4{{#1} {\bf #2}, #3 (#4)}

\def\JHEP{\em JHEP}

\def\PLB{{\em Phys. Lett.}  B}

\def\PRD{{\em Phys. Rev.} D}

\def\ra{\rightarrow}

\def\be{\begin{equation}}
\def\ee{\end{equation}}
\def\bea{\begin{eqnarray}}
\def\eea{\end{eqnarray}}

\begin{document}
\vspace*{4cm}
\title{SUPERSYMMETRIC DARK MATTER -- A PHYSICS GODOT?\footnote{Invited
talk at the Rencontres de Moriond ``ElectroWeak Interactions and
Unified Theories'', 9--16 March, 2002}}

\author{ LESZEK ROSZKOWSKI}

\address{Department of Physics, Lancaster University\\
        Lancaster LA1 4YB, England}

\maketitle\abstracts{Where is the long--awaited one? A supersymmetric
neutralino has been a favored candidate for the WIMP dark matter but, so far,
it has not been found. One way to locate it is to identify where it can
be hiding in the vast supersymmetric parameter space. A combination of
theoretical, experimental and cosmological constraints leads to 
remarkably well--defined allowed regions which favor lighter
neutralino, and other superpartner, mass ranges. Nevertheless the neutralino
can still be as heavy as roughly $800\gev$. The resulting scalar
direct detection cross sections range
from some $10^{-7}$~pb down to $10^{-10}$~pb, which is at least one
order of magnitude below the sensitivity of today's experiments but
which will be almost fully explored by a new generation of planned detectors. 
}

\section{Why SUSY and Why WIMPs}\label{sec:why}

You may wonder why I have pointed a finger on supersymmetric dark matter
(SUSY DM) as a guest who has failed to show up, the ``Godot'' that we have
been waiting and waiting for so long, and so far with no luck. After
all, one can list many more ``Godot's'' in physics, so why pick on
SUSY DM. Without looking too far, 
no unambiguous signal of SUSY itself has been found yet, after
some twenty years of searching. Nor have we managed to find out what
the real nature of the dark matter in the Universe is since some seventy
years ago when the DM problem was first identified in the Coma cluster
by Zwicky. 

We do not know for sure what the SUSY DM is but we have a rather good
idea what it is likely to be. The main suspect is of course the LSP,
the lightest SUSY particle, which is stable due to an assumed
$R$--parity.  Simulations of large structure formation and CMB studies
point towards cold DM.  Astrophysical constraints tell us that such a
particle should not be electrically charged, nor (preferably) should
it interact strongly, hence another acronym has been coined for the
favorite class of DM: a WIMP, the weakly interacting massive
particle. SUSY's LSP has long been known to fit the bill pretty
well. In fact, if anything, the list could only be shorter.  A
commonly considered candidate for the CDM is the lightest
neutralino. The neutralino is known to often provide a sizeable
contribution to the relic density for reasonable ranges of other
superpartner masses. But the list does not end there.  The axino
(a superpartner of the axion)~\cite{ckr} and the gravitino are also very
well-motivated and attractive. The only problem with them would be
that, as ``Godots'', one would have to wait for them even longer due
to their exceedingly faint interactions with ordinary matter, means
detectors.

In this talk I will focus on the neutralino ``Godot''. I will try to
see where this guest who has failed to show up, may be hiding. The
SUSY parameter space is known to be large. It is still rather mildly
constrained from below by collider searches, while from above
superpartner masses are not expected to exceed a few~\tev\ due to a
somewhat vague `naturalness criterion'. Fortunately, much more
constraining information is provided by non-collider searches,
especially $BR(B\rightarrow X_s\gamma)$ and the anomalous magnetic moment of
the muon $(g-2)_{\mu}$, as well as
cosmological considerations: the age of the Universe and direct
determinations of the WIMP relic abundance $\abundchi$. I will present
here results of a recent comprehensive analysis. Due to the
lack of space, I will quote here only the papers presented
here - extensive sets of references can be found there.

\section{Constrained MSSM}\label{sec:cmssm}
The framework I will adopt is that of the Constrained MSSM (CMSSM), as
defined in the original paper where the acronym was
introduced~\cite{kkrw94} and what most theorists mean by it. (This is in
contrast to what LEP experimentalists would refer to as CMSSM today, or rather,
yesterday.)

In the [theorist's] CMSSM, in addition to the
requirement of a common gaugino mass $\mhalf$ at the unification scale
$\mgut$, which is usually made in the more generic Minimal
Supersymmetric Standard Model (MSSM), one further assumes that the
soft masses of all scalars (sfermion and Higgs) are equal to $\mzero$
at $\mgut$, and analogously that the trilinear soft terms unify at
$\mgut$ at some common value $\azero$. These parameters are run using
their respective Renormalization Group Equations (RGEs) from $\mgut$
to some appropriately chosen low-energy scale $\qzero$ where the Higgs
potential (including full one-loop corrections) is minimized, while
keeping the usual ratio $\tanb$ of the Higgs VEVs fixed. The
Higgs/higgsino mass parameter $\mu$ and the bilinear soft mass term $\bmu$
are next computed from the conditions of radiative electroweak
symmetry breaking (EWSB), and so are the Higgs and superpartner masses. The
CMSSM thus has a priori only the usual 
$\tanb,\mhalf, \mzero, \azero, \sgn(\mu) $
as input parameters. 
However, in the case of large $\mhalf,\mzero\gsim1\tev$ and/or large
$\tanb\sim {\cal O}(\mt/\mb) $ some resulting masses will in general
be highly sensitive to the assumed physical masses of the top and the
bottom (as well as the tau), and they will also
strongly depend on the correct choice of the scale $\qzero$. This in
particular will affect the impact of the cosmological constraints
as I will discuss below.

In the CMSSM, the LSP neutralino is often a nearly pure
bino~\cite{rr93,kkrw94} because the requirement of radiative
EWSB typically gives $|\mu|\gg \mone$ where $\mone$ is the soft mass
of the bino.  This often (albeit not
always~\cite{kkrw94}) allows one to impose strong constraints from
$\abundchi< {\cal O}(1)$ on $\mhalf$ and $\mzero$ (and therefore also
on heaviest Higgs and superpartner masses) in the ballpark of $1\tev$.
This was originally shown in Refs.~\cite{rr93,kkrw94} and later
confirmed by many subsequent studies.

\section{Mass Spectra and Experimental Constraints}\label{sec:expt}

In the case of the CMSSM, the most important experimental constraints
from LEP are those on the masses of the lightest chargino
$\mcharone>104\gev$ and Higgs boson $\hl$. For a Standard Molel-like
Higgs, the bound is $\mhl>113.5\gev$ but one should keep in mind that,
due to large radiative corrections, the theoretical uncertainty in
$\mhl$ in the CMSSM is probably of the order of $2$--$3\gev$. I will
therefore show the Higgs mass contours corresponding to the value
given above and also to $111\gev$ as a more conservative value.

Let's next turn to non-accelerator constraints.  First, there has been
much activity in determining $\br( B\ra X_s \gamma )$. A recent
combined experimental result, which incorporates the new CLEO result,
gives $\br( B\ra X_s \gamma ) =(3.23\pm0.42)\times 10^{-4}$.  This
allows for some, but not much, room for contributions from SUSY when
one compares it with the updated prediction for the Standard Model
(SM) $\br( B\ra X_s \gamma ) = (3.73\pm 0.30)\times 10^{-4}$. Second,
at large $\tanb$ next-to-leading order supersymmetric corrections to
$b\ra s\gamma$ become important.  In our analysis we adopt the full
expressions for the dominant terms.  We add the two $1\sigma$ errors
(the experimental and SM) in quadrature and further add linearly $0.2$
to accommodate the theoretical uncertainty in SUSY contributions which
is roughly $5\%$ of the SM value for branching ratio. Altogether we
conservatively allow our results to be in the range $\br( B\ra X_s
\gamma ) = (3.23\pm 0.72)\times 10^{-4}$ for SM plus two-Higgs
doublets plus superpartner contribution. The excluded regions of SUSY
masses will not however be extremely sensitive to the choice of these
error bars but instead to the underlying assumption of minimal flavor
violation.~\cite{or1}

The first measurement
by the Brookhaven experiment E821 of the anomalous magnetic moment of
the muon last year gave 
$\amu=(g_\mu-2)/2$. After some ups and
downs with correcting sign errors in theory calculations, the
result implies a mild $1.6\sigma$ discrepancy between the experimental
value and the SM prediction
$\amuexpt-\amusm=(25.6\pm16.6)\times10^{-10}$. As we will see, this will
provide an upper limit on the plane of $(\mhalf,\mzero)$ at
the $1\sigma$ level (also disfavoring negative $\mu$), but it will
quickly evaporate if one takes a slighly more conservative approach.

The details of our procedure for obtaining the mass spectra can be found in
Ref.~\cite{rrn1}. 
We calculate superpartner and Higgs 
mass spectra using the two-loop RGEs for the gauge, Yukawa and soft
mass parameters. Appropriate QCD corrections are included which become
important especially at large $\tanb$. 
Of particular importance is a correct treatment of the Higgs sector and
the conditions for the EWSB. We include full one-loop corrections to 
the Higgs potential and minimize it at
$\qstop\sim\sqrt{\mstopone \mstoptwo}$. 
The mass of the pseudoscalar will play a crucial role in computing
$\abundchi$, especially at very large $\tanb\sim50$.
This is so for three reasons: 
$\mha$ becomes now much smaller than at
smaller $\tanb$ due to the increased role of the bottom Yukawa coupling;
because the $A$-resonance in $\chi\chi\ra f\bar{f}$ is
dominant since the coupling $Af\bar{f}\sim \tanb$ for down-type
fermions; and because, in
contrast to the heavy scalar $\hh$, this channel is not $p$-wave
suppressed. In ISASUGRA $\mha$ is computed as
$\mha^2= \left(\tanb + \cot\beta\right)\left( -\bmu + \Delta_A^2\right)$
where $\Delta_A^2$ stands for 
the full one-loop corrections which can be
significant.

\section{Results}\label{sec:results}

In Figs.~1a and~b I present two typical distint cases. The first
applies to values of $\tanb$ up to around 45, the second to larger
values. All the grey, red and light orange regions are excluded as
described in the captions. In particular, the last one
($\abundchi>0.3$) comes from the lower limit on the age of the
Universe and clearly provides an extremely impressive constraint. The
narrow white bands are allowed by cosmology ($\abundchi<0.3$) while
the [even narrower] green strips correspond to the favored range
$0.1<\abundchi<0.2$. The relic abundance $\abundchi$ can now be
evaluated at a per cent level.~\cite{nrr1,nrr2}

It is clear that in the left window only two very thin regions are
allowed. The horizontal region at $\mzero\sim$ few hundred~\gev\ is, at
lower $\mhalf$, excluded by the chargino and Higgs mass bounds at
smaller $\tanb$ and by the $BR(B\rightarrow X_s\gamma)$ constraint at
large $\tanb$. In fact, were it not for the coannihilation of the
neutralino with the lighter stau $\stauone$, much of it (on the right
side) would be excluded by $\abundchi>0.3$.
The one at $\mzero\gg\mhalf$ is disfavored by the current
value of $(g-2)_{\mu}$ and, to some extent, by naturalness arguments. 

The visible difference between the two windows comes from the very
wide pseudoscalar Higgs resonance in the annihilation process
$\chi\chi\rightarrow A\rightarrow f {\bar f}$. Since $m_A$
decreases with increasing $\tanb$, at some point, this opens up a
corridor in the plane of ($\protect{\mhalf},\protect{\mzero}$) 
along $m_A=2\mchi$.

The regions consistent with all the theoretical, experimental and
cosmological constraints are indeed remarkably small, when compared to
the whole available parameter space, especially at
$\tanb\lsim45$. This has clear implications for the Higgs and
superpartner masses and other properties. Unfortunately, the ``Godot''
we are after, the (bino-like) neutralino LSP, while confined to the
allowed regions, is not all that well-constrained. Its mass
$\mchi\sim0.4\mhalf$ can still
be as large as some $800\gev$ (at very large $\tanb\gsim50$), or even
more, but only
because of the coannihilation with the stau and because of the wide
$A$--resonance. The resulting direct detection WIMP-proton cross
sections, a characteristic quantity for DM WIMP searches, ranges from
some $10^{-7}$~pb down to $10^{-10}$~pb, or so, but I have no time to
explain it.

\begin{figure}[t!]
\vspace*{-0.1in}
\hspace*{-.12in}
\begin{minipage}{8in}
\epsfig{file=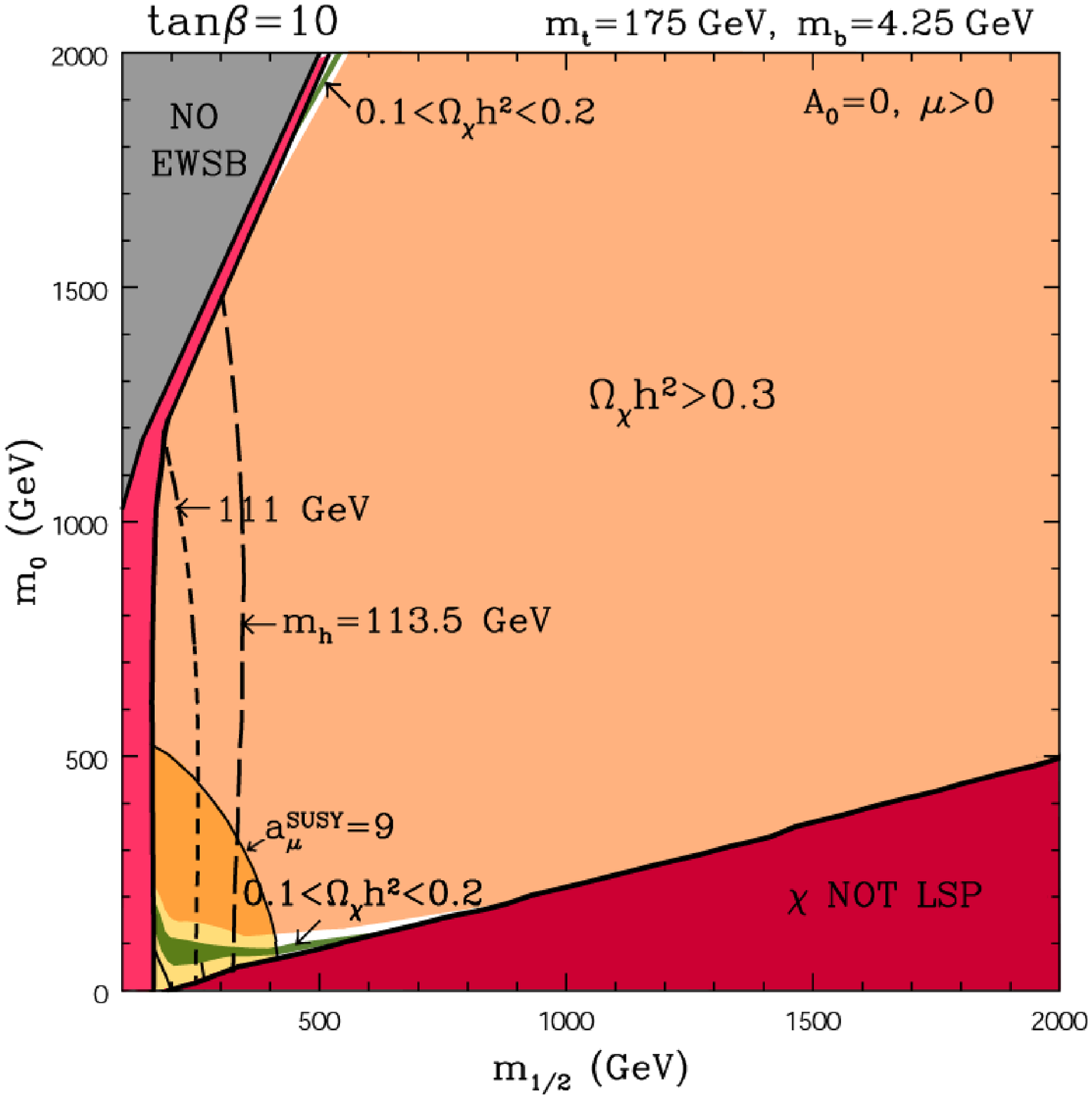,height=3.15in} 
\hspace*{-0.03in}
\epsfig{file=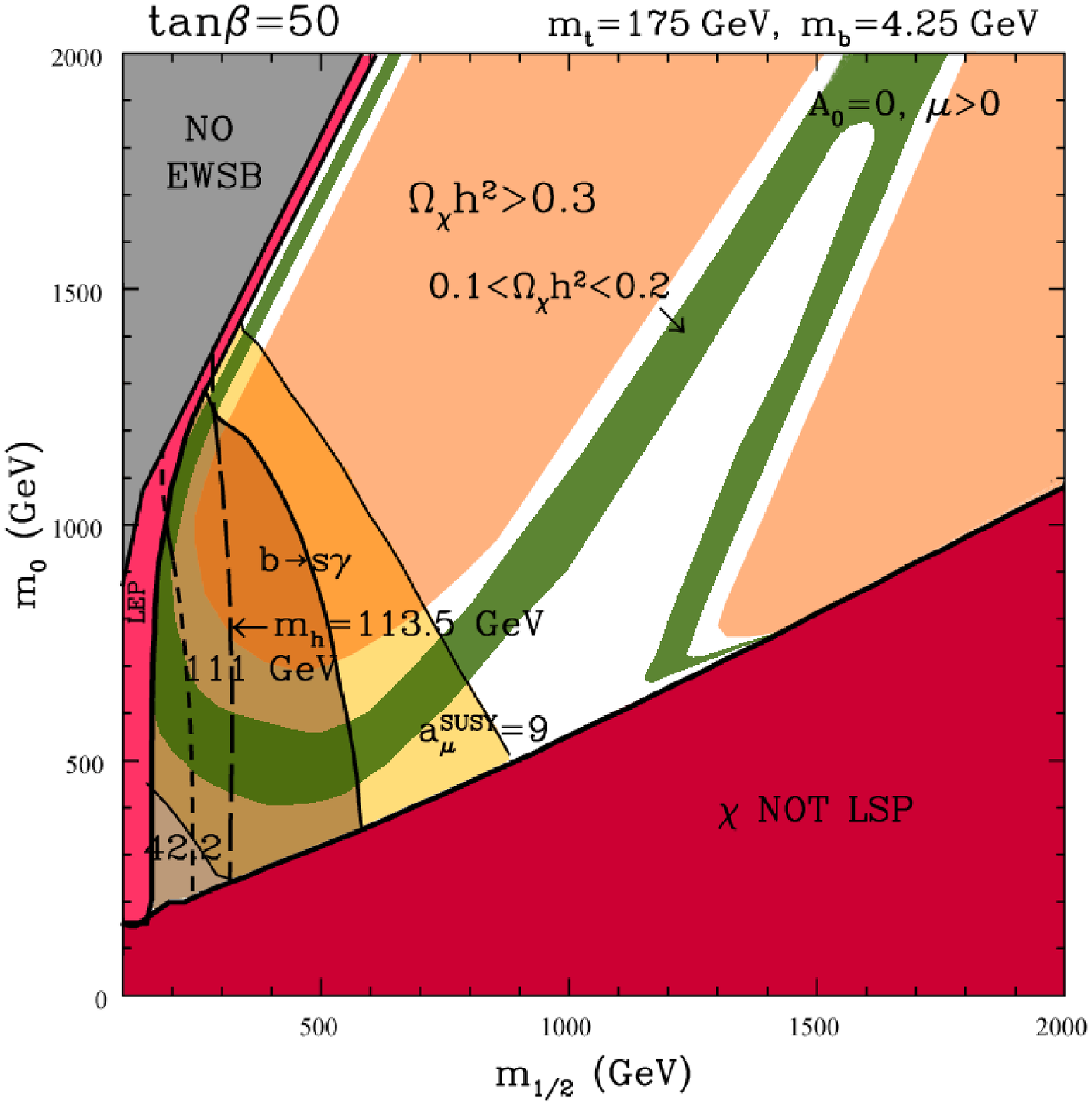,height=3.15in} 
\end{minipage}
\caption{\label{fig:tb50detail} The plane 
($\protect{\mhalf},\protect{\mzero}$) for $\protect{\tanb=10}$ (left window),
$\protect{\tanb=50}$ (right window) and for 
$\azero=0$, $\mu>0$, $\mt\equiv\mtpole=175\gev$ and
$\mb\equiv\mbmbsmmsbar=4.25\gev$. The light red bands on the left are
excluded by chargino searches at LEP. In the grey wedge in the
left-hand corner electroweak symmetry breaking conditions are not
satisfied. The dark red region denoted `$\chi$ NOT LSP' corresponds to
the lighter stau being the LSP. The large light orange regions of
$\abundchi>0.3$ are excluded by cosmology while the narrow green bands
correspond to the expected range $0.1<\abundchi<0.2$. The light brown
region in the right window is excluded by the
lower bound of $\br(B\ra X_s \gamma ) = (3.23\pm 0.72)\times 10^{-4}$. 
Also shown are
the semi-oval dark yellow contours of $\amususy\equiv\deltaamususy/10^{-10}$
favored by the anomalous magnetic moment of the muon measurement at
$1\sigma$~CL
($\amususy=9,42.2$). The lines of the lightest
Higgs scalar mass $\mhl=111\gev$ and $113.5\gev$ are denoted by short
and long-dash lines, respectively.}
\end{figure}

\vspace*{0.3cm}
So, instead of waiting for the SUSY DM ``Godot'' the strategy would be to
go and get him. We theorists have pointed a finger on where he may be
hiding. It is now up to our experimentalist colleagues to do the rest
of the job. This is what the next talk is going to address.


\end{document}